\documentclass[12pt]{article}
\usepackage{amsmath,amssymb,amsthm,color}
\usepackage{graphicx}
\usepackage{cite}
\usepackage{epsfig}
\renewcommand{\baselinestretch}{2}
\begin{document}
%
\title{Chemical Bondings Induced Rich Electronic Properties of Oxygen Absorbed Few-layer Graphenes\\}
\author{
\small Ngoc Thanh Thuy Tran$^{a}$, Shih-Yang Lin$^{a}$, Yu-Tsung Lin$^{a}$, Ming-Fa Lin$^{a,*}$ $$\\
\small $^a$Department of Physics, National Cheng Kung University, Tainan 701, Taiwan \\
 }
\renewcommand{\baselinestretch}{1}
\maketitle

\renewcommand{\baselinestretch}{1.4}
\begin{abstract}

Electronic properties of graphene oxides enriched by the strong chemical bondings are investigated using first-principle calculations. They are very sensitive to the changes in the number of graphene layer, stacking configuration, and distribution of oxygen. The feature-rich electronic structures exhibit the destruction  or distortion of Dirac cone, opening of band gap, anisotropic energy dispersions, O- and (C,O)-dominated energy dispersions, and extra critical points. All the few-layer graphene oxides are semi-metals except for the semiconducting monolayer ones. For the former, the distorted Dirac-cone structures and the O-dominated energy bands near the Fermi level are revealed simultaneously. The orbital-projected density of states (DOS) have many special structures mainly coming from a composite energy band, the parabolic and partially flat ones. The DOS and spatial charge distributions clearly indicate the critical bondings in O-O, C-O and C-C bonds, being responsible for the diversified properties. 
\vskip 1.0 truecm
\par\noindent

\vskip1.0 truecm

\par\noindent  * Corresponding authors. {~ Tel:~ +886-6-2757575-65272 (M.F. Lin)}\\~{{\it E-mail address}: mflin@mail.ncku.edu.tw (M.F. Lin), sylin.1985@gmail.com (S.Y. Lin)}
\end{abstract}
\pagebreak
\renewcommand{\baselinestretch}{2}
\newpage

{\bf 1. Introduction}
\vskip 0.3 truecm

Graphene has been a mainstream material in science and engineering for so many years, mainly owing to its remarkable physical, chemical and material properties. Numerous researchers have tried to diversify the essential properties of graphene by using doping \cite{liu2011chemical,wei2009synthesis,nakada2011migration}, stacking configuration \cite{zhong2012stacking,tran2015configuration}, layer number \cite{sutter2009electronic,lauffer2008atomic}, electric or/and magnetic fields \cite{lai2008magnetoelectronic,huang2014feature,lu2006influence,tang2011electric}, and mechanical strain \cite{wong2012strain,pereira2009strain}. Monolayer graphene has a low-energy isotropic Dirac-cone structure, while it is a zero-gap semiconductor because of the vanishing density of states (DOS) at the Fermi level ($E_F$=0). On the other hand, all the few-layer graphenes are semi-metals with small overlaps in valence and conduction bands, in which the low-lying energy bands are significantly affected by the layer number and stacking configurations. Recently, oxygen absorbed graphene as so-called graphene oxide (GO) has been the focus of attention due to its opening of band gap and other interesting properties \cite{dikin2007preparation,mkhoyan2009atomic,hirata2004thin}. It has promised potential applications in wide variety of areas, including memristor devices \cite{Porro2015383}, sensors \cite{veerapandian2012synthesis,robinson2008reduced}, and supercapacitors \cite{Xue2015305,chen2011high,gao2011direct}. The paper seeks to contribute to critical orbital hybridizations in O-O, C-O and C-C bonds by investigating how oxygen absorption, layer number and stacking configuration can enrich the electronic properties of graphene.\\

The well known method to produce GO is the Hummers method in which a mixture of sodium nitrate, sulphuric acid and potassium permanganate are used in the synthesis process \cite{hummers1958preparation}. Recently, other methods have been developed by adding phosphoric acid combined with the sulphuric acid without sodium nitrate, and increasing the amount of potassium permanganate \cite{marcano2010improved}. The oxygen concentration of 70\% can be reached in this modified method \cite{marcano2010improved}. The distribution and concentration of oxygen can be examined using nuclear magnetic resonance (NMR) \cite{cai2008synthesis}, and X-ray photoelectron spectroscopy (XPS) \cite{gao2009new}, respectively. On the other hand, theoretical studies have shown the dependence of band gap on the concentration and distribution of oxygen atoms in monolayer graphene oxide \cite{lian2013big,ito2008semiconducting}. The oxygen absoption has converted a part of sp$^2$ carbon bonds in pristine graphene to sp$^3$ bonds in GO \cite{mkhoyan2009atomic}. It has been reported that the strong hybridization between the 2$p_{x,y}$ orbitals of O atoms and the 2$p_z$ orbitals of C atoms is responsible for the opening of band gap \cite{lian2013big}. However, this work shows that the above-mentioned orbital hybridization is not exact (discussed later in Figs. 4 and 5 ). The important orbital hybridizations in O-O, C-O and C-C bonds are not fully analyzed, and the oxygen absorbed few-layer graphenes system has not been investigated. Therefore, the critical roles of such bondings in determining the essential properties of oxygen absorbed few-layer graphenes are worthy of a systematic study.\\

In this paper, the strong effects of oxygen atoms on the electronic properties of graphene, with different distributions, layers, and stackings, are investigated using first-principles calculations. The calculated results demonstrate that the bond lengths, energy dispersions, band gap, charge distributions, and density of states (DOS) are sensitive to the changes in oxygen distributions and geometric structures. Specifically, the energy bands exhibit a lot of features: the destruction or distortion of Dirac cone, opening of band gap, partially flat bands, energy dispersions related to O-O and C-O bonds, and extra critical points. Such features are reflected in the orbital-projected DOS, including the absence and presence of $\pi$ band peaks, the O-dominated special structures near $E_F$, and the (C,O)-dominated prominent peaks at middle energy. Furthermore, the diverse orbital hybridizations in C-O, O-O, and C-C bonds will be explored through the DOS and spatial charge distributions to explain the dramatic changes in electronic properties. The above-mentioned results can be verified by experimental measurements such as ARPES \cite{ohta2006controlling}, and scanning tunneling spectroscopy (STS) \cite{lauffer2008atomic}.\\

\vskip 0.6 truecm
\par\noindent
{\bf 2. Methods }
\vskip 0.3 truecm

The first-principles calculations on GO are performed based on density functional theory (DFT) using the Vienna Ab initio Simulation Package (VASP) \cite{kresse1996efficient,kresse1999ultrasoft}. The electron-ion interactions are evaluated by the projector augmented wave method \cite{blochl1994projector}, whereas the electron-electron interactions are taken into account by the exchange-correlation function under the generalized gradient approximation of Perdew-Burke-Ernzerhof \cite{perdew1996generalized}. A vacuum layer with a thickness of 12 $\mbox\AA$ is added in a direction perpendicular to the GO plane to avoid interactions between adjacent unit cells. The wave functions are expanded using a plane-wave basis set with a maximum kinetic energy of 500 eV. All atomic coordinates are relaxed until the Hellmann-Feynman force on each atom is less than 0.01 eV/$\mbox\AA$. Furthermore, the van der Waals (vdW) force is employed in the calculations using the semiemprical DFT-D2 correction of Grimme to correctly describe the atomic interactions between graphene layers \cite{grimme2006semiempirical}. The k-point mesh is set as 200$\times$200$\times$1 in geometry optimization, 30$\times$30$\times$1 in band structure, and 250$\times$250$\times$1 in the DOS for the 1$\times$1 unit cell with zigzag structure. An equivalent k-point mesh is set for other cells depending on their size.\\

\vskip 0.6 truecm
\par\noindent
{\bf 3. Results and discussion}
\vskip 0.3 truecm

\begin{figure}[htb]
\centering\includegraphics[width=6cm]{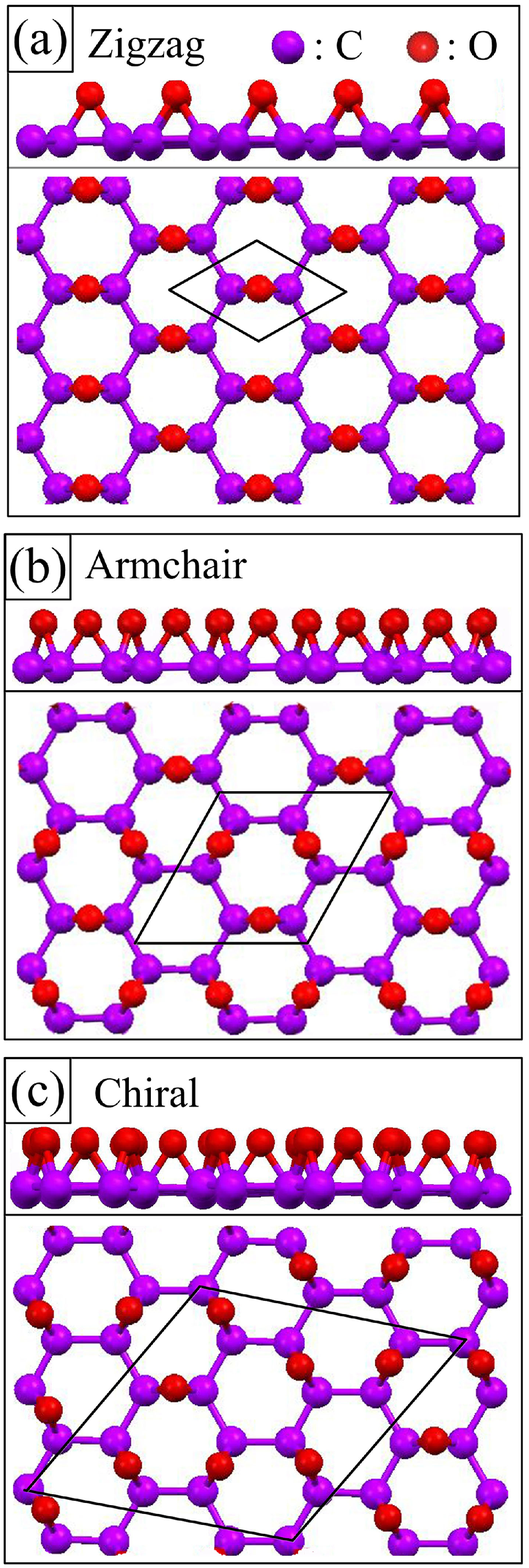}
\caption{Geometric structures of 50\% oxygen concentration systems with: (a) zigzag, (b) armchair, (c) chiral unit cells.}
\end{figure}

 The atomic structures of oxygen absorbed monolayer graphene with top view and side view are shown in Fig. 1. Oxygen atoms are absorbed on the top graphene layer with O/C ratio of 1/2 (50\%) for zigzag (Fig. 1(a)), armchair (Fig. 1(b)), and chiral (Fig. 1(c)) unit cells, in which carbon atoms are arranged along these edge structures, respectively. In comparison with the two latters, the first one has a lower total ground state energy meaning more stable. This is in agreement with previous studies which have shown that the higher the symmetric of the geometric structure, the more stable the system is \cite{huang2012oxygen}. However, the difference in total ground state energy among these configurations is only about 7-8 \%, indicating that all the distributions are considerably stable. From the top view, oxygen atoms are at the top between two carbon atoms as so-called the bridge-site. We have checked the most stable sites when oxygen is absorbed on graphene by comparing their total ground state energies. Among the bridge-, hollow- and top-sites, the bridge-site is the most stable one and the hollow-site is the less stable one, which agrees with previous studies \cite{nakada2011migration,saxena2011investigation}. \\
 
\begin{table}[htb]
 \caption{The calculated C-O bond lengths, and C-C bond lengths of the zigzag, armchair, and chiral unit cell with various layers.}
 \label{t1}
 \begin{center}
 \begin{tabular}{ c c c c c c c c }
 \hline
 Graphene Oxides & C-O bond & Nearest C-C bond & Second nearest C-C \\
 & length (${\AA}$)& length (${\AA}$)& bond length (${\AA}$)\\
 \hline
 Monolayer (zigzag)& 1.434 & 1.503 & 1.543\\
 Bilayer & 1.436 & 1.459 & 1.491\\
 Trilayer & 1.437 & 1.441 & 1.470\\
 Four-layer & 1.438 & 1.432 & 1.459\\
 Five-layer & 1.438 & 1.431 & 1.453\\
 Six-layer & 1.438 & 1.430 & 1.450\\
 Seven-layer & 1.438 & 1.423 & 1.449\\
 Eight-layer & 1.439 & 1.418 & 1.444\\
 Nice-layer & 1.439 & 1.416 & 1.444\\
 \hline
 Monolayer (armchair) & 1.410 & 1.543 & 1.569\\
 Bilayer (armchair) & 1.413 & 1.477 & 1.497\\
 \hline
 Monolayer (chiral) & 1.424 & 1.519 & 1.560\\
 Bilayer (chiral) & 1.429 & 1.465 & 1.496\\
 \hline
 \end{tabular}
 \end{center}
 \end{table}
 
 The main features of geometric structures, including the C-C bond lengths, C-O bond lengths, and interlayer distances are dominated by the number of graphene layer (n). As oxygen atoms are absorbed on graphene, the C-C bond lengths are expanded compared to 1.424 $\mbox\AA$ in pristine graphene, as shown in Table 1. The equivalent C-C bond lengths are altered after oxidation, in which the nearest C-C bond corresponding to oxygen atom on the bridge side is shorter than the second nearest one, as a result of the strong C-O bond (Fig. 4(b)). The longer C-C bond length in GO means the 2$p_{x,y}$ orbitals have the hybridization with atomic orbitals of O. As the layer number grows n$\ge$2, the C-C bond lengths show a significant decrease while the C-O bond length only has a slight rise. They remain almost unchanged and close to that of pristine graphene for n$\ge$8. Our calculation also reveals that the interlayer distances are affected by oxygen absorption. The optimized values for AA and AB stackings are equal to 3.27 $\mbox\AA$ instead of 3.52 $\mbox\AA$ and 3.26 $\mbox\AA$ in pristine bilayer systems \cite{tran2015configuration}. Similarly, for AAA and ABA stackings, the interlayer distance between the middle and top layers is about 3.27 $\mbox\AA$, whereas that of the bottom and middle layers are nearly identical to the AA- and AB-bilayer ones, respectively. Further examinations, it is found that the n-layer graphene with oxygen absorbed on the top can be qualitatively regarded as the (n$-$1)-layer of graphene with monolayer graphene oxide. The chemical bonding of atomic orbitals will be investigated in more detail later (Fig. 4). \\
 
 \begin{figure}[htb]
 \centering\includegraphics[width=10cm]{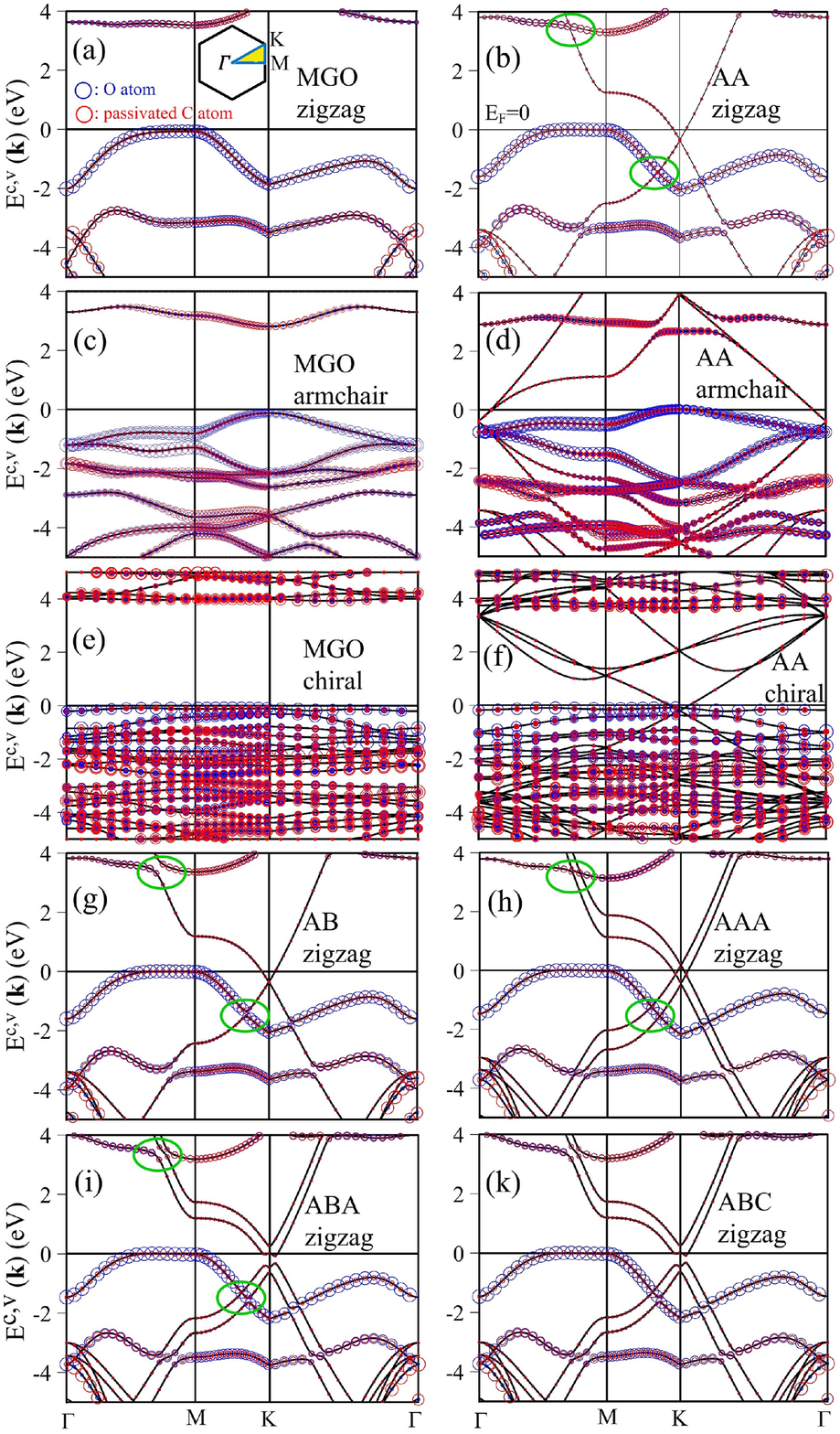}
 \caption{Band structures of the oxygen-absorbed few-layer graphenes: (a) monolayer (zigzag), (b) AA bilayer (zigzag), (c) monolayer (armchair), (d) AA bilayer (armchair), (e) AB bilayer (zigzag), (f) AAA trilayer (zigzag), (g) ABA trilayer (zigzag), and (h) ABC trilayer (zigzag). Superscripts "c" and "v" correspond to the conduction and valence band, respectively. Also shown in the inset of (a) is the first Brillouin zone.}
  \end{figure}
 
 The two-dimensional (2D) energy bands along high symmetric points are determined by the C-O, O-O and C-C bonds, oxygen distribution, layer number and stacking configuration (Fig. 2). The contributions of the O atoms and the C atoms passivated with the former are represented by the blue and red circles, respectively, in which the dominance is proportional to the circle's radius. Different from monolayer graphene, the isotropic Dirac-cone structure near the K point is destroyed in GO, mainly owing to the serious hybridization between the atomic orbitals of C and O atoms (discussed later in charge density and DOS). Such strong C-O bonds lead to the termination of the complete $\pi$ bonds between parallel $2p_z$ orbitals of C atoms accounting for the Dirac-cone structure near $E_F$ in pristine graphene. Instead, there is a wide direct gap of 3.5 eV, 2.8 eV, and 3.9 eV for the zigzag, armchair, and chiral unit cells, as shown in Figs. 2(a), 2(c) and 2(e), respectively. The highest occupied state is related to an O-dominated energy band. The O-O bond has the weaker $\sigma$ bond which arises from the (2$p_y-$2$p_y$), (2$p_{x,y}-$2$p_{x,y}$) and (2$p_{x,y}-$2$p_{x,y}$) hybridizations for zigzag, armchair and chiral distributions, respectively (Figs. 4 and 5). As a result, the O-dominated low-lying bands might have the partially flat or parabolic dispersions for the different distributions. On the other hand, the above-mentioned C-O bonds induce the (C,O)-related energy bands at -2 eV$\le\,E^v\le$ -4 eV because of the strong orbital hybridizations between passivated C and O atoms (blue and red circles, revealed simultaneously). As to the planar C-C bond, it possesses the strongest $\sigma$ bond formed by the (2$p_x$, 2$p_y$, 2s) orbitals and thus creates the deeper $\sigma$ bands with $E^v\le\,-3.5$ eV (red circles).\\
 
 For bilayer GO, the band gap is replaced by a distorted Dirac-cone structure that comes from the C
 atoms on another layer without O passivation. The Dirac point is revealed at the K point for zigzag (Fig. 2(b)) and chiral unit cells (Fig. 2(f)) and at the $\Gamma$ point for armchair one (Fig. 2(d)). The anisotropic linear bands will change into the parabolic bands with the saddle point along K $\rightarrow$ M or $\Gamma\rightarrow$ M. For example, the saddle points of the $\pi$ and $\pi^*$ bands, respectively, correspond to $E^v\approx-2.5$ eV and $E^c\approx1.3$ eV for the zigzag  distribution in Fig. 2(b). Different from AA bilayer graphene in which two isotropic Dirac-cone structures with the linear bands intersecting at $E_F$ \cite{tran2015configuration}, there is one anisotropic Dirac-cone structure with the Fermi level located at the conduction band. There still exist the O-dominated energy dispersions near $E_F$ and the (C,O)-related bands at middle energy, as revealed in monolayer GO. The band-edge states of the former are above the Fermi level, indicating its free-hole density equal to the free-electron one in the distorted Dirac cone. As the number of graphene layer increases, there are more distorted Dirac cones and saddle points. The band structure differences between AA and AB stackings (Fig. 2(g)) are small and only lie in the crossing and anti-crossing bands (green ellipses). Specially, the energy dispersions of ABA and ABC stackings (Fig. 2(k)) are almost identical, since these two systems correspond to the same structure of AB bilayer graphene with monolayer GO on the top, as proposed ealier in the geometric structures. However, these two systems are different from the AAA stacking in the distorted Dirac-cone structures as well as the crossing and anti-crossing bands (Figs. 2(h) and 2(i)). In addition, the Fermi velocities of the distorted Dirac cones are reduced about 7-10\% compared with that of pristine graphene ($\approx 10^6$ m/s) \cite{neto2009electronic}.\\ 
 
 \begin{figure}[htb]
 \centering\includegraphics[width=14cm]{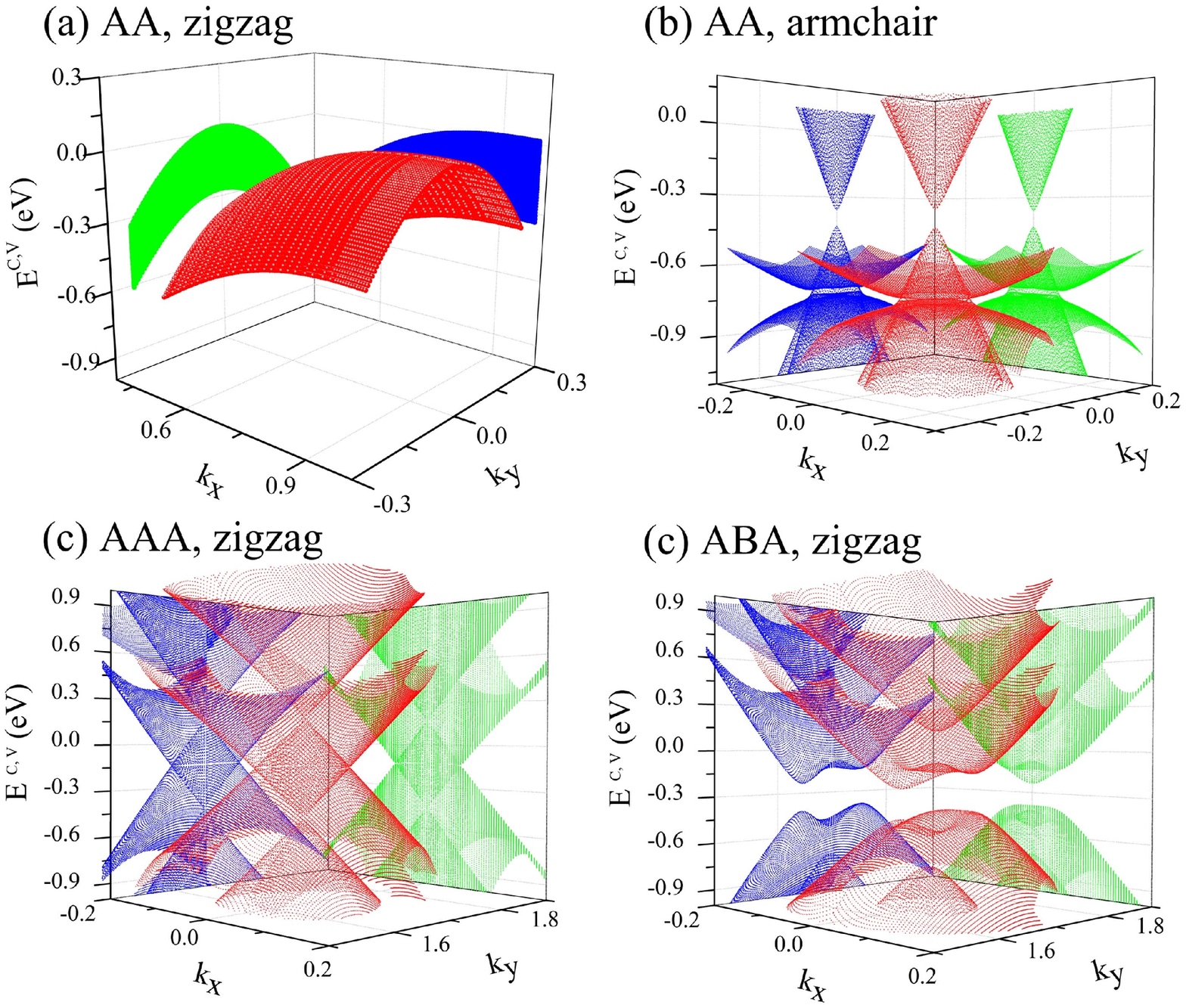}
 \caption{Low-lying 3D band structures for: (a) zigzag AA bilayer near the M point, (b) armchair AA bilayer near the Gamma point, (c) zigzag AAA trilayer near the K point, and (d) zigzag ABA trilayer near the K point.}
 \end{figure}
 
 The critical low-energy electronic properties can be further understood by the three-dimensional (3D) bands shown in Fig. 3. The zigzag AA stacking, respectively, exhibits the partially flat and parabolic dispersions on the xz and yz planes (blue and green regions) with the crossing of the Fermi level. Therefore, the O-dominated energy band can be regarded as a 1D parabolic band. However, this band has the parabolic (the partially flat dispersions) on the planar projections for the armchair (chiral) distribution, as shown in Fig. 2(c) (Fig. 2(e)). As to the distorted Dirac-cone structures, their main features include the change in energy dispersion, the separation of Dirac points, and the crossing with the O-dominated energy band. For example, the armchair AA stacking has an energy spacing of 50 meV between two separate Dirac points (near the $\Gamma$ point in Fig. 3(b)) and one pair of O-dominated parabolic bands at $E^v\simeq\,-0.75$ eV crossing the valence Dirac cone (Fig. 2(d)). The low-lying energy bands of zigzag AAA and ABA stackings, as shown in Figs. 3(c) and 3(d), clearly illustrate the significant differences in energy dispersions and spacings. The former are two pairs of linear bands with $\sim\,10$ meV energy spacing, whereas the later consist of one pair of parabolic bands and one pair of Mexican-hat bands with $\sim\,0.1$ eV energy spacing.\\
 
 Recently, ARPES has emerged as a most useful experimental technique to study the electronic band structures. The experimental measurements on graphene-related systems have been used to investigate the effects due to doping \cite{zhou2008metal}, layer number \cite{ohta2007interlayer,sutter2009electronic}, stacking configuration \cite{ohta2006controlling}, and the electric field \cite{ohta2006controlling}. For example, they have verified the Dirac-cone structure of graphene grown on SiC \cite{ohta2008morphology}, and observed the opening of band gap for graphene on Ir(111) through oxidation \cite{schulte2013bandgap}. As expected, the feature-rich energy bands of oxygen absorbed few-layer graphenes, including the absence and presence of the distorted Dirac-cone structures, the band gap, the O-dominated bands near $E_F$, and the (C,O)-dominated bands at middle energy can be examined by ARPES. The comparisons between theoretical predictions and experimental measurements can comprehend how oxygen distribution, layer number and stacking configuration affect the electronic properties of oxygen absorbed few-layer graphenes.\\
 
 The charge density ($\rho$) and the charge density difference ($\Delta$$\rho$) can provide very useful information on the spatial charge redistributions and thus on the dramatic changes of energy bands \cite{lin2015feature}. The former reveals the chemical bondings as well as the charge transfer. In pristine AA stacking (Fig. 4(a)), a strong covalent $\sigma$ bond with high charge density due to the (2$p_x$,2$p_y$,2s) orbitals exists between two C atoms (the pink rectangle; details in Fig. 4(d)). Furthermore, there is a weak $\pi$ bond arising from the parallel 2$p_z$ orbitals at the boundary (the purple rectangle). When O atoms are absorbed on zigzag AA stacking surface (Fig. 4(b)), charges transferred from C to O atoms are $\sim\,1e$ using the Bader charge analysis (the dashed black rectangle). The strong C-O bond, respectively, induces the termination of $\pi$ bond on the upper boundary and the charge redistribution of $\pi$ bond on another one. The former accounts for the absence of Dirac cone and the opening of energy gap. The latter is further reflected in the $\pi$ bond of the bottom layer by the vdW interactions, so that the Dirac cone presents the distorted structure (Fig. 2(b)). The above-mentioned characteristics about the $\sigma$ and $\pi$ bonds as well as the strong C-O bond are also found in armchair and chiral distributions (not shown).  \\
 
 \begin{figure}[htb]
 \centering\includegraphics[width=14cm]{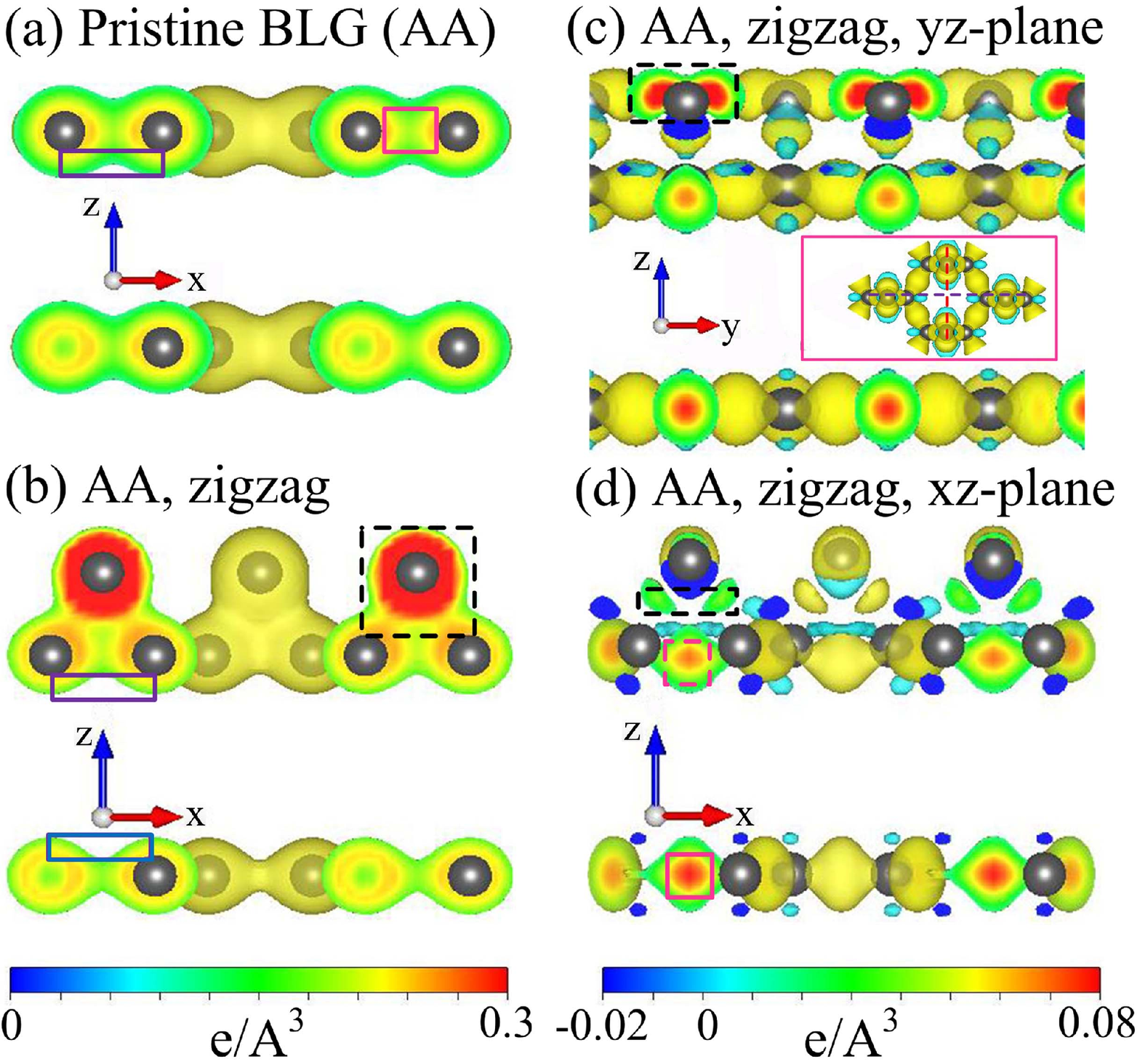}
 \caption{The charge density $\rho$ of (a) pristine AA stacking, and (b) zizag AA bilayer. The charge density difference $\Delta\rho$ of zigzag AA bilayer in (c) yz-plane, and (d) xz-plane. }
  \end{figure}
 
 In order to comprehend the orbital hybridizations in O-O, C-O and C-C bonds, the variation of charge density of zigzag AA bilayer is illustrated in Figs. 4(c) and 4(d). $\Delta$$\rho$ is created by substracting the charge density of isolated C and O atoms from that of GO system. The top view of charge density difference of zigzag AA stacking is shown in the inset of Fig. 4(c), where the red and purple dashed lines represent the slices in yz and xz planes, respectively. The $p_y$ orbitals of O with high charge density enclosed in the dashed black rectangle (Fig. 4(c)) are lengthened along $\hat{y}$ compared to isolated O atom's, clearly indicating the 2$p_y-$2$p_y$ hybridization. The O-O bond belongs to a weak $\sigma$ bond because of the larger distance, being the main reason for the low-lying O-related bands near $E_F$ (Fig. 2(b)). It has also been observed that the O-O bond comes from the hybridization of 2$p_{x,y}-$2$p_{x,y}$ for the armchair and chiral distributions (not shown: Figs. 5(c) and 5(e)). In sharp contrast to the 2$p_y$ orbitals, the 2$p_z$ and 2$p_x$ orbitals of O have strong hybridizations with those of C, as seen from the green region enclosed by the dashed purple rectangle (Fig. 4(d)). This region lies between O and passivated C atoms and bents toward the latter, while the $p_y$ orbitals do not reveal any hybridization with other orbitals. In fact, it is hard to completely distinguish which orbitals have hybridizations to each other by using charge density analysis, but they also show the significant evidence in the orbital-projected DOS (Fig. 5). The C-O bond is much stronger than the O-O bond; therefore, the (C,O)-related bands appear at the range of -2 eV$\le\,E^v\le$-4 eV. On the other hand, between two non-passivated C atoms of GO, $\Delta$$\rho$ shows a strong $\sigma$ bond indicated by the enclosed pink square in Figs. 4(d). Such bond becomes a bit weaker after the C-O bond is formed (the dashed pink square in Figs. 4(d)). This demonstrates that not only the 2$p_z$ orbitals of passivated C atoms hybridize with orbitals of O atoms but also the 2$p_{x,y}$ orbitals play an important role. However, the planar C-C bond remains the strongest one and has the deeper $\sigma$ energy bands.\\
 
  The main characteristics of band structures are directly reflected in DOS. The orbital-projected DOS, as shown in Fig. 5, is useful in understanding the orbital contributions as well as the orbital hybridizations in chemical bonds. The feature-rich DOS presents three kinds of special structures, including the asymmetric peak (circle), the shoulder structures, and the symmetric peaks. They, respectively, correspond to a composite band with partially flat and parabolic dispersions, the minimum/maximum band-edge states of parabolic bands, and the saddle points of parabolic bands or the partially flat bands \cite{wong2012strain}. The low-energy DOS is thoroughly altered after oxygen absorption. For pristine graphene, the peaks caused by $\pi$ and $\pi^*$ bands due to 2$p_z$-2$p_z$ bondings between C atoms will dominate within the range of $|E^{c ,v}|\le$ 2 eV \cite{neto2009electronic}. However, the $\pi$- and $\pi^*$-peaks are absent as a result of the strong C-O bond. Instead, there are several O-dominated prominent structures in a wide range of $E\sim$-2.5 eV to $E_F$, mainly coming from the 2$p_y-$2$p_y$ bondings in zigzag distribution (Fig. 5(a)) and the 2$p_{x,y}-$2$p_{x,y}$ bondings in armchair and chiral ones (Figs. 5(c) and 5(e)). It is noticed that the contributions of 2$p_x$ and 2$p_y$ orbitals in armchair distribution are equivalent because of the symmetric distribution of oxygen. With the addition of graphene layer, the $\pi$- and $\pi^*$-peaks related to the non-passivated C atoms appear around -2.5 eV and 1.2 eV, respectively (arrows in Figs. 5(b), 5(d); 5(f)-5(h)). Apparently, the number of these peaks are only affected by graphene layer number (Figs. 5(b), 5(g) and 5(h)). \\
 
 \begin{figure}[htb]
 \centering\includegraphics[width=14cm]{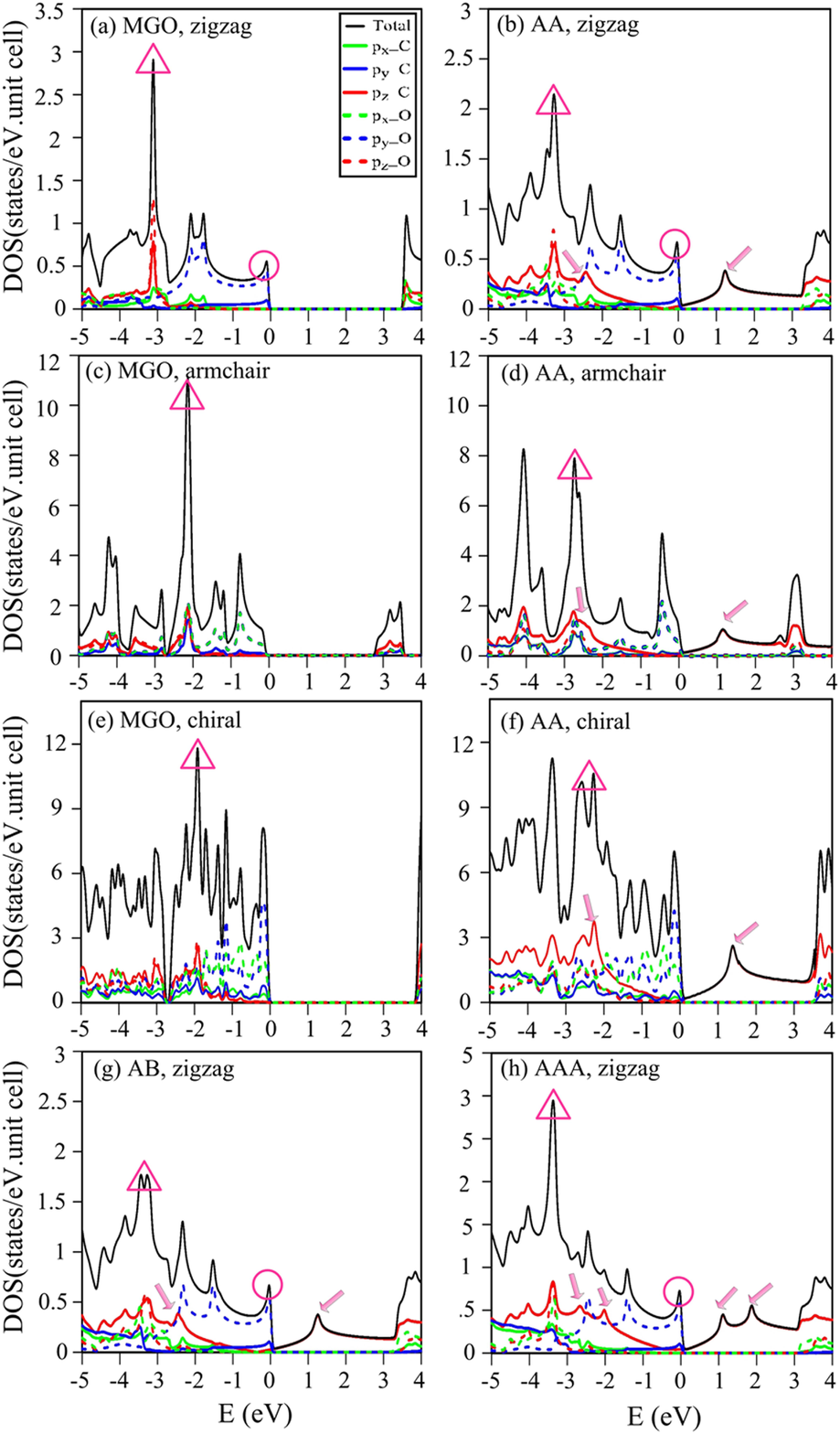}
 \caption{DOS of the oxygen-absorbed few-layer graphenes: (a) zigzag monolayer, (b) zigzag AA bilayer, (c) armchair AA bilayer, (d) chiral AA bilayer, (e) zigzag AB bilayer, and (f) zigzag AAA trilayer.}
  \end{figure}
 
 With the increasing energy, DOS grows quickly and exhibit prominent symmetric peaks due to the C-O bond at middle energy (-2 eV$\le\,E^v\le$-4 eV). Among these peaks, the most pronounced one is mainly contributed by the (2$p_z$,2$p_x$), (2$p_x$,2$p_y$,2$p_z$) and (2$p_x$,2$p_y$,2$p_z$) orbitals of C and O atoms for the zigzag, armchair and chiral distributions, respectively (e.g., triangles in Figs. 5(a), 5(c); 5(e)). The contributions of distinct orbitals revealed at the same energy apparently represent the significant hybridizations between them, being consistent with those in the spatial charge distributions (Fig. 4(d)). This strong covalent C-O bond is the main reason for the absence of $\pi$ and $\pi^*$ peaks and the opening of band gap. Also, the magnitude of energy gap depends on the O-O bond strength, as indicated from distinct gaps among three types of  distributions.\\
  
 Up to now, there are several theoretical studied on GO, especially in electronic properties \cite{lian2013big,ito2008semiconducting,mkhoyan2009atomic,sutar2012electronic}. They are focused on how the band gap is modulated with the variation of oxygen concentrations \cite{lian2013big,ito2008semiconducting}. The orbital hybridizations that cause the opening of band gap and critical electronic properties have not been fully explored. In this work, the comprehensive orbital hybridizations in C-O, O-O, and C-C bonds are analyzed by the spatial charge distributions and orbital-projected DOS (Figs. 4 and 5). The previous study \cite{lian2013big} proposed 2$p_z-$2$p_{x,y}$ orbital hybridizations of C and O atoms to explain the opening of band gap, in which they are not totally exact. Our results indicate orbital hybridizations of 2p$_{x,z}-$2$p_{x,z}$ or 2p$_{x,y,z}-$2$p_{x,y,z}$ between passivated C and O atoms, obviously including the 2$p_z$-2$p_z$ hybridizations. The effect of oxygen distributions on the magnitude of band gap has been pointed out \cite{lian2013big,mkhoyan2009atomic}, but not the chemical picture for this, the 2$p_y-$2$p_y$ or 2$p_{x,y}-$2$p_{x,y}$ hybridizations in O-O bond. Moreover, the terminated $\pi$ bond, the reformed $\pi$ and $\sigma$ bonds are clearly illustrated using spatial charge distributions (Fig. 4).\\

 The STS measurement, in which the tunneling conductance (dI/dV) is approximately proportional to DOS and directly reflects its special structures, can provide an efficient way to confirm the distribution of oxygen. This method has been used for adatoms on graphene \cite{gyamfi2011fe}, carbon nanotubes \cite{wilder1998electronic}, graphene nanoribbons \cite{huang2012spatially}, and few-layer graphene \cite{choi2010atomic}. For example, it has been used to observe the Fermi-level shift and additional peaks near Dirac point of graphene irradiated with Ar$^+$ ions \cite{tapaszto2008tuning}. The main features in electronic properties, including the energy gaps, the O-dominated prominent structures near $E_F$, the high $\pi$- and $\pi^*$-peaks, and the strong (C,O)-related peaks at middle energy, can be further investigated with STS. The STS measurements on the low- and middle-energy peaks can identify the complex chemical bondings in oxygen absorbed few-layer graphenes.\\ 

\vskip 0.6 truecm
\par\noindent
{\bf 4. Conclusion }
\vskip 0.3 truecm


In summary, the geometric structures and electronic properties of oxygen absorbed few-layer graphenes are studied using first-principles calculations. They are shown to be dominated by the diverse chemical bondings between atomic orbitals. The C-C bond lengths are expanded due to oxygen absorption and gradually recover to that in pristine graphene with the increasing layer number. The nearest C-C bond length is shorter compared to the second nearest one, indicating the significant interactions between the passivated C atoms and O atoms. The C-C, C-O and O-O bonds are responsible for the dramatic changes in electronic properties, including the destruction or distortion of Dirac cone, energy gap, anisotropic energy spectra, (C,O)- and O-dominated energy bands, and many extra critical points. The oxygen absorbed few-layer graphenes, with the O-related energy bands and the distorted Dirac cone near $E_F$, are semi-metals except for the semiconducting monolayer ones. The predicted rich-feature band structures, which depend on the oxygen distribution, staking configuration and layer number, could be examined by the ARPES measurements.\\

The competition or cooperation among the critical chemical bondings in C-C, O-O and C-O bonds can enrich the essential properties. The complete $\pi$ bondings due to the parallel 2$p_z$ orbitals in C-C bonds can be formed on the graphene planes without oxygen adsorption. They are affected by the oxygen passivation so that the distorted Dirac-cone structures exhibit the crossing with the O-dominated band, the energy spacings between the separated Dirac points, and the changes in energy dispersions. Such structures are sensitive to the number of layers and stacking configurations, e.g., the same bands between ABA and ABC stackings, but the distinct bands between ABA and AAA stackings. In addition, the (2$p_x$,2$p_y$,2s) orbitals of C atoms only create the $\sigma$ bands with the deeper state energies lower than $-$3.5 eV. The 2$p_y-$2$p_y$ or 2$p_{x,y}-$2$p_{x,y}$ orbital hybridizations in the O-O bonds are determined by the oxygen distribution; furthermore, the sufficiently strong bondings will generate the O-dominated energy dispersions with a bandwidth of $\sim$2.5 eV near $E_F$. There exist serious orbital hybridizations of 2p$_{x,z}-$2$p_{x,z}$ or 2p$_{x,y,z}-$2$p_{x,y,z}$ between passivated C and O atoms, thus, leading to the deeper (C,O)-related energy bands at -2 eV$\le\,E^v\le$-4 eV, and the absence of the isotropic Dirac cone, which has not been sufficiently explored in previous studies \cite{lian2013big,mkhoyan2009atomic}. The main features of the orbital-dependent energy bands are directly reflected in a lot of peaks and shoulder structures in DOS. The experimental examinations of STS on DOS, as well as those of ARPES on energy bands, can provide the full information to understand the complex orbital bondings in graphene oxides.

\par\noindent {\bf Acknowledgments}

This work was supported by the Physics Division, National Center for Theoretical Sciences (South), the Nation Science Council of Taiwan (Grant No. NSC 102-2112-M-006-007-MY3). We also thank the National Center for High-performance Computing (NCHC) for computer facilities.

\newpage
\renewcommand{\baselinestretch}{0.2}

\newpage \centerline {\Large \textbf {FIGURE CAPTIONS}}

\vskip0.5 truecm 

 Fig. 1. Geometric structures of 50\% oxygen concentration systems with: (a) zigzag, (b) armchair, and (c) chiral unit cells.

 Fig. 2. Band structures of the oxygen-absorbed few-layer graphenes: (a) monolayer (zigzag), (b) AA bilayer (zigzag), (c) monolayer (armchair), (d) AA bilayer (armchair), (e) AB bilayer (zigzag), (f) AAA trilayer (zigzag), (g) ABA trilayer (zigzag), and (h) ABC trilayer (zigzag). Superscripts $c$ and $v$ correspond to the conduction and valence bands, respectively. Also shown in the inset of (a) is the first Brillouin zone.

 Fig. 3. Low-lying 3D band structures for: (a) zigzag AA bilayer near the M point, (b) armchair AA bilayer near the Gamma point, (c) zigzag AAA trilayer near the K point, and (d) zigzag ABA trilayer near the K point.
 
 Fig. 4.  The charge density $\rho$ of (a) pristine AA stacking, and (b) zizag AA bilayer. The charge density difference $\Delta\rho$ of zigzag AA bilayer in (c) yz-plane, and (d) xz-plane. 
 
 Fig. 5. DOS of the oxygen-absorbed few-layer graphenes: (a) zigzag monolayer, (b) zigzag AA bilayer, (c) armchair AA bilayer, (d) chiral AA bilayer, (e) zigzag AB bilayer, and (f) zigzag AAA trilayer.


\begin{thebibliography}{99}
\bibitem{liu2011chemical} Liu HT, Liu YQ, Zhu DB. Chemical doping of graphene. J. Mater. Chem. 2011, 21(10): 3335-3345.
%
\bibitem{wei2009synthesis} Wei DC, Liu YQ, Wang Y, Zhang HL, Huang LP, Yu G. Synthesis of N-doped graphene by chemical vapor deposition and its electrical properties. Nano Lett. 2009, 9(5):1752-1758.
%
\bibitem{nakada2011migration} Nakada K, Ishii A. Migration of adatom adsorption on graphene using DFT calculation. Solid State Commun. 2011, 151(1): 13-16.
%
\bibitem{zhong2012stacking} Zhong X, Pandey R, Karna SP. Stacking dependent electronic structure and transport in bilayer graphene nanoribbons. Carbon 2012, 50(3): 784-790.

\bibitem{tran2015configuration} Tran NTT, Lin SY, Glukhova OE, Lin MF. Configuration-induced rich electronic properties of bilayer graphene. J. Phys. Chem. C 2015, 119: 10623-10630.

\bibitem{sutter2009electronic} Sutter P, Hybertsen MS, Sadowski JT, Sutter E. Electronic structure of few-layer epitaxial graphene on Ru (0001). Nano Lett. 2009, 9(7): 2654-2660.

\bibitem{lauffer2008atomic} Lauffer P, Emtsev KV, Graupner R, Seyller T, Ley L, Reshanov SA, Weber HB. Atomic and electronic structure of few-layer graphene on SiC (0001) studied with scanning tunneling microscopy and spectroscopy. Phys. Rev. B 2008, 77(15):155426. 

\bibitem{lai2008magnetoelectronic} Lai YH, Ho JH, Chang CP, Lin MF. Magnetoelectronic properties of bilayer Bernal graphene. Phys. Rev. B 2008, 77(8):085426.

\bibitem{huang2014feature} Huang YK, Chen SC, Ho, YH, Lin CY, Lin MF. Feature-rich magnetic quantization in sliding bilayer graphenes. Sci. Rep. 2014, 4: 7509.

%
\bibitem{lu2006influence} Lu CL, Chang CP, Huang YC, Chen RB, Lin ML. Influence of an electric field on the optical properties of few-layer graphene with AB stacking. Phys. Rev. B 2006, 73(14): 144427.

\bibitem{tang2011electric} Tang KC, Qin R, Zhou J, Qu, H, Zheng JX, Fei RX, Li H, Zheng, QY, Gao ZX, Lu J. Electric-field-induced energy gap in few-layer graphene. J. Phys. Chem. C 2011, 115(19): 9458-9464.
%
\bibitem{wong2012strain} Wong JH, Wu BR, Lin MF. Strain effect on the electronic properties of single layer and bilayer graphene. J. Phys. Chem. C 2012, 116(14): 8271-8277.
%
\bibitem{pereira2009strain} Pereira VM, Neto AHC. Strain engineering of graphene's electronic structure. Phys. Rev. Lett. 2009, 103(4): 046801.
%
\bibitem{dikin2007preparation} Dikin DA, Stankovich S, Zimney EJ, Piner RD, Dommett GHB, Evmenenko G, Nguyen SBT, Ruoff RS. Preparation and characterization of graphene oxide paper. Nature 2007, 448(7152): 457-460.
%
\bibitem{mkhoyan2009atomic} Mkhoyan KA, Contryman AW, Silcox J, Stewart DA, Eda G, Mattevi C, Miller S, Chhowalla M. Atomic and electronic structure of graphene-oxide. Nano Lett. 2009, 9(3): 1058-1063.

\bibitem{hirata2004thin} Hirata M, Gotou T, Horiuchi S, Fujiwara M, Ohba M. Thin-film particles of graphite oxide 1: High-yield synthesis and flexibility of the particles. Carbon 2004, 42(14): 2929-2937.

\bibitem{Porro2015383} Samuele P, Eugenio A, Candido FP, Carlo R. Memristive devices based on graphene oxide. Carbon 2015, 85: 383-396.
%

\bibitem{veerapandian2012synthesis} Veerapandian M, Lee MH, Krishnamoorthy K, Yun K. Synthesis, characterization and electrochemical properties of functionalized graphene oxide. Carbon 2012, 50(11): 4228-4238.

\bibitem{robinson2008reduced} Robinson JT, Perkins FK, Snow ES, Wei Z, Sheehan PE. Reduced graphene oxide molecular sensors. Nano Lett. 2008, 8(10): 3137-3140.
%
\bibitem{Xue2015305} Xue YH, Zhu L, Chen H, Qu J, Dai LM. Multiscale patterning of graphene oxide and reduced graphene oxide for flexible supercapacitors. Carbon 2015, 92: 305-310.

\bibitem{chen2011high} Chen Y, Zhang X, Zhang D, Yu P, Ma Y. High performance supercapacitors based on reduced graphene oxide in aqueous and ionic liquid electrolytes. Carbon 2011, 49(2): 573-580.

\bibitem{gao2011direct} Gao W, Singh N, Song L, Liu Z, Reddy ALM, Ci L, Vajtai R, Zhang Q, Wei B, Ajayan PM. Direct laser writing of micro-supercapacitors on hydrated graphite oxide films. Nat. Nanotechnol. 2011, 6(8): 496-500.
%
\bibitem{hummers1958preparation} Hummers J, William S and Offeman RE. Preparation of graphitic oxide. J. Am. Chem. Soc. 1958, 80(6), 1339-1339.

\bibitem{marcano2010improved} Marcano DC, Kosynkin DV, Berlin JM, Sinitskii A, Sun Z, SlesarevA, Alemany LB, Lu W, Tour JM. Improved synthesis of graphene oxide. ACS Nano 2010, 4(8):4806-4814.

\bibitem{cai2008synthesis} Cai W, Piner RD, Stadermann FJ, Park S, Shaibat MA, Ishii Y, Yang D, Velamakanni A, An SJ, Stoller M et. al. Synthesis and solid-state NMR structural characterization of 13C-labeled graphite oxide. Science 2008, 321(5897): 1815-1817.

\bibitem{gao2009new} Gao W, Alemany LB, Ci L, Ajayan PM. New insights into the structure and reduction of graphite oxide. Nat. Chem. 2009, 1(5): 403-408.

\bibitem{lian2013big} Lian KY, Ji YF, Li XF, Jin MX, Ding DJ, Luo Y. Big bandgap in highly reduced graphene oxides. J. Phys. Chem. C 2013, 117(12): 6049-6054.

\bibitem{ito2008semiconducting} Ito, Jun and Nakamura, Jun and Natori, Akiko. Semiconducting nature of the oxygen-adsorbed graphene sheet. J. App. Phys. 2008, 103(11): 113712.

\bibitem{ohta2006controlling} Ohta T, Bostwick A, Seyller T, Horn K, Rotenberg E. Controlling the electronic structure of bilayer graphene. Science 2006, 313(5789): 951-954.
%

\bibitem{kresse1996efficient} Kresse G, Furthm{\"u}ller J. Efficient iterative schemes for ab initio total-energy calculations using a plane-wave basis set. Phys. Rev. B 1996, 54(16): 11169.
%

\bibitem{kresse1999ultrasoft} Kresse G, Joubert D. From ultrasoft pseudopotentials to the projector augmented-wave method. Phys. Rev. B 1999, 59(3): 1758.

\bibitem{blochl1994projector} Bl{\"o}chl PE. Projector augmented-wave method. Phys. Rev. B 1994, 50(24): 17953.

\bibitem{perdew1996generalized} Perdew JP, Burke K, Ernzerhof M. Generalized gradient approximation made simple. Phys. Rev. Lett. 1996, 77(18): 3865.

\bibitem{grimme2006semiempirical} Grimme S. Semiempirical GGA-type density functional constructed with a long-range dispersion correction. J. Comput. Chem. 2006, 27(15): 1787-1799.

\bibitem{huang2012oxygen} Huang H, Li Z, She J, Wang W. Oxygen density dependent band gap of reduced graphene oxide. J. App. Phys. 2012, 111(5): 054317.

\bibitem{saxena2011investigation} Saxena S, Tyson TA, Shukla S, Negusse E, Chen H, Bai J. Investigation of structural and electronic properties of graphene oxide. App. Phys. Let. 2011, 99(1): 013104.

\bibitem{neto2009electronic} Neto AHC, Guinea F, Peres NMR, Novoselov KS, Geim AK. The electronic properties of graphene. Rev. Mod. Phys. 2009, 81(1): 109.

\bibitem{zhou2008metal} Zhou SY, Siegel DA, Fedorov AV, Lanzara A. Metal to insulator transition in epitaxial graphene induced by molecular doping. Phys. Rev. Lett. 2008, 101(8):086402.

\bibitem{ohta2007interlayer} Ohta T, Bostwick A, McChesney JL, Seyller T, Horn K, Rotenberg E. Interlayer interaction and electronic screening in multilayer graphene investigated with angle-resolved photoemission spectroscopy. Phys. Rev. Lett. 2007, 98(20): 206802.

\bibitem{ohta2008morphology} Ohta T, El Gabaly F, Bostwick A, McChesney JL, Emtsev KV, SchmidAK, Seyller T, Horn K, Rotenberg E. Morphology of graphene thin film growth on SiC (0001). New J. Phys. 2008, 10(2): 023034.
%
\bibitem{schulte2013bandgap} Schulte K, Vinogradov NA, Ng ML, M{\aa}rtensson N, Preobrajenski AB. Bandgap formation in graphene on Ir (111) through oxidation. App. Surf. Sci. 2013, 267: 74-76.

\bibitem{lin2015feature} Lin SY, Chang SL, Shyu FL, Lu JM, Lin MF. Feature-rich electronic properties in graphene ripples. Carbon 2015, 86: 207-216.

\bibitem{sutar2012electronic} Sutar DS, Singh G, Botcha VD. Electronic structure of graphene oxide and reduced graphene oxide monolayers. App. Phys. Lett. 2012, 101(10): 103103. 

\bibitem{gyamfi2011fe} Gyamfi M, Eelbo T, Wa{\'s}niowska M, Wiesendanger R. Fe adatoms on graphene/Ru (0001): Adsorption site and local electronic properties. Phys. Rev. B 2011, 84(11): 113403.

\bibitem{wilder1998electronic} Wilder JWG, Venema LC, Rinzler AG, Smalley RE, Dekker C. Electronic structure of atomically resolved carbon nanotubes. Nature 1998, 391{6662}:59-62.

\bibitem{huang2012spatially} Huang H, Wei D, Sun J, Wong SL, Feng YP, Neto AHC, Wee ATS. Spatially resolved electronic structures of atomically precise armchair graphene nanoribbons. Sci. Rep. 2012, 2: 983.

\bibitem{choi2010atomic} Choi J, Lee H, Kim S. Atomic-scale investigation of epitaxial graphene grown on 6H-SiC (0001) using scanning tunneling microscopy and spectroscopy. J. Phys. Chem. C 2010, 114(31): 13344-13348.


\bibitem{tapaszto2008tuning} Tapaszto L, Dobrik G, Nemes IP, Vertesy G, Lambin P, Bir{\'o} LP. Tuning the electronic structure of graphene by ion irradiation. Phys. Rev. B 2008, 78(23): 233407.

\end{thebibliography}
\end{document}